# Um Estudo Sobre Arquitetura de Redes Neurais Aplicado à Previsão do Retorno de Ações Brasileiras

EnANPAD - 2017

Leonardo Felizardo – Fundação Getúlio Vargas - leonardo.felizardo@usp.br

Afonso Pinto – Fundação Getúlio Vargas - afonso.pinto@fgv.br

**Resumo**

Neste trabalho é apresentada uma análise estatística sobre as características que entendemos influenciar no desempenho das redes neurais em termos de assertividade na previsão de retornos de ações brasileiras. Criamos uma população de arquiteturas para análise e desta extraímos a amostra que teve o melhor desempenho assertivo. Verificou-se como as características desta amostra se destacam e afetam as redes neurais. Além disso, fazemos inferências com relação à que tipo de influência as diferentes arquiteturas têm no desempenho das redes neurais. No estudo é realizado a previsão do retorno de uma ação brasileira negociada na bolsa de valores de São Paulo para mensurar o erro cometido pelas diferentes arquiteturas de redes neurais construídas. Os resultados são promissores e indicam que alguns aspectos da arquitetura de redes neurais têm impacto significativo na assertividade do modelo.

**Introdução**

Redes neurais, também conhecidas como redes neurais artificiais, trata-se de uma tecnologia originada de muitas disciplinas: neurociência, matemática, estatística, física, ciência da computação e engenharia. Sua principal virtude é a habilidade de aprender a partir de dados de entrada, com ou sem supervisão (HAYAKIN, 2008). Em diversos trabalhos é possível observar as redes neurais sendo comparadas a métodos tradicionais do mundo das finanças, tais como análise técnica, análises fundamentalistas, séries temporais, análise de auto correlação, entre outras. Em particular, no trabalho de Behrammirzaee (2010) são mencionados diversos estudos com a comparação das redes neurais com outros métodos de finanças. Nota-se que as redes neurais têm um desempenho variado, até pela própria característica dos problemas apresentados, embora em muitos casos tenham se mostrado mais adequadas do que outras abordagens. Há bastante espaço para o desenvolvimento e para otimizar na sua utilização. Um questionamento que pode ser feito é: de fato a melhor rede está sendo configurada para aquele determinado problema? Boa parte dos trabalhos apresentados não dão argumentações fortes com relação à metodologia utilizada para a seleção da arquitetura da rede neural aplicada ao problema em questão. É possível verificar que embora em muitos casos os autores expliquem o racional para a escolha das variáveis das redes, não trabalham para otimizar a arquitetura para determinado objetivo. Isso é observável em diversos estudos tais como o de Faria (2009) que, embora realize alguns testes, não foca na otimização da arquitetura, que pode ter impacto tanto na precisão como no tempo de processamento, aspecto este muito relevante para operações de alta frequência. De fato, assertividade e tempo de processamento são altamente impactados pela arquitetura de rede neural. Embora existam várias formas de otimizar fatores que compõem a arquitetura de redes neurais, ainda há uma carência na literatura por metodologias consistentes que expliquem o impacto das características inerentes destas redes e explicações de como elas afetam e otimizam todas as variáveis envolvidas no problema. Neste trabalho, seguimos como abordagem a análise de um grande número de arquiteturas pelo teste exaustivo de todas as redes possíveis, dadas as condições de contorno limitantes necessárias para o seu processamento. O principal objetivo do trabalho é, portanto, prover percepções e considerações para que futuras aplicações de redes neurais em finanças tenham uma boa base de referência para a escolha e seleção de suas arquiteturas, além de uma visão técnica mais sistemática com relação à definição das variáveis de arquitetura das redes neurais consideradas. O estudo aqui proposto tem a seguinte estrutura e conteúdo. O item 2 apresenta uma breve revisão bibliográfica a respeito do histórico das redes neurais, suas aplicações e otimização de arquiteturas. O item 3 faz uma revisão dos conceitos teóricos aqui utilizados. O item 4 detalha a metodologia utilizada para a realização dos experimentos. O item 5 apresenta os resultados obtidos a partir dos experimentos realizados e, finalmente, o item 6 apresenta as conclusões tiradas a partir dos resultados, assim como indica algumas limitações do trabalho e sugestões para pesquisas futuras.

## 2 Revisão Bibliográfica

Após alguns importantes trabalhos que consolidaram redes neurais como uma poderosa ferramenta de previsão, o trabalho de Webor (1974) supera as limitações computacionais do *perceptron* de camada única proposto por Rosenblatt (1958), com a introdução de conceitos que envolvem a utilização de *perceptrons* de múltiplas camadas treinados com algoritmo de retropropagação, A partir deste momento, diversas aplicações surgiram utilizando-se das redes neurais artificiais. Da mesma forma que se desenvolveram outras técnicas estatísticas, suas primeiras aplicações foram em automação e controle, em meados dos anos 1990 começaram a surgir algumas aplicações na área de finanças. Destas, destacamos o trabalho de McNelis (1996) que utiliza ferramentas de classificação de padrões, *Data Mining* e previsão. As aplicações de redes neurais artificiais são variadas e divididas em diversas áreas. Behrammirzaee (2010) divide em sua pesquisa as principais aplicações realizadas: (i) avaliação de crédit (ABDOU, POINTON e EL-MASRY, 2008); (ii) planejamento financeiro e previsão, (MARIJANA e KLIME POPOSKI, 2009); (iii) gestão de portfólio (ZIMMERMANN, 2001); (iv) Em boa parte dos trabalhos destacados por Behrammirzaee, é possível perceber que as redes neurais propostas desempenharam melhor do que os métodos mais convencionais utilizados para fins comparativos. No Brasil, destacamos alguns trabalhos ligados à precificação de opções no mercado brasileiro (CHAGAS, 2006), à avaliação de crédito bancário em instituição financeira brasileira (ARNAUD, CUNHA, *et al.*, 2005), à recuperação de receita com prevenção de insolvência (PINHEIRO, EVSUKOFF e EBECKEN, 2006), à variação de cotação de ações (ZANETI e ALMEIDA, 1998), e à previsão de rentabilidade de empresas (MATSUMOTO, 2008). Nestes estudos, existem diversas comparações entre as redes neurais e outros modelos, tais como: modelos lineares (MEDEIROS, VEIGA e PEDREIRA, 2001), *K-Means* (ARNAUD, CUNHA, *et al.*, 2005) e Black &Scholes (FREITAS, 2001). Conforme as aplicações das redes neurais se expandiram, a importância de desenvolver arquiteturas mais eficientes cresceu. Alguns trabalhos tratam o tema de definição de arquitetura com mais profundidade, utilizando-se de técnicas tais como: (i) algoritmos genéticos (BENARDOS e VOSNIAKOS, 2006); (ii) método Taguchi (KHAW, LIM e LIM, 1994); (iii) métodos construtivos (BALKIN e ORD, 2000); (iv) entre outros que tratam de forma genérica a otimização de redes neurais. Estes trabalhos, em sua maioria não estão ligados à área de finanças, tendo suas aplicações essencialmente em engenharia e automação.

Este estudo busca fornecer a partir da análise das características das arquiteturas de redes neurais, insumos para escolha de uma arquitetura ótima para o problema de previsão de retorno de ações. A metodologia aqui aplicada ainda pode ser expandida para outros problemas fora do mundo de finanças, fornecendo assim uma visão mais assertiva de como adequadamente escolher uma arquitetura de rede neural.

# 3 Redes Neurais

A construção de redes neurais artificiais é inspirada na forma como o cérebro humano realiza o seu processamento, que é consideravelmente diferente da forma como os computadores o fazem. O cérebro tem um processamento altamente complexo, não-linear e paralelo, dando-lhe condições de realiza r o processamento de determinadas tarefas de forma muito mais eficiente do que um computador. De forma geral, uma rede neural artificial é uma máquina projetada com o objetivo de simular a maneira como o cérebro opera. Os componentes que compõem este modelo podem ser eletrônicos ou algoritmos de programação, e seguem a mesma estrutura básica de uma rede neural orgânica, ou seja, empregam interligações maciças de células computacionais denominadas "neurônios" ou "unidades de processamento". A definição formal adotada por Hayakin (2008) é: "uma rede neural é um processador maciço e paralelamente distribuído, constituído de unidades de processamento simples, que têm a propensão natural para armazenar conhecimento experimental e torná-lo disponível para o uso". Tentamos modelar o neurônio orgânico com artifícios em programação criando unidades de processamento de informações como ilustrado na Figura 1. Estas unidades também são conhecidas como nós, pois são os pontos de cruzamento das diferentes conexões.

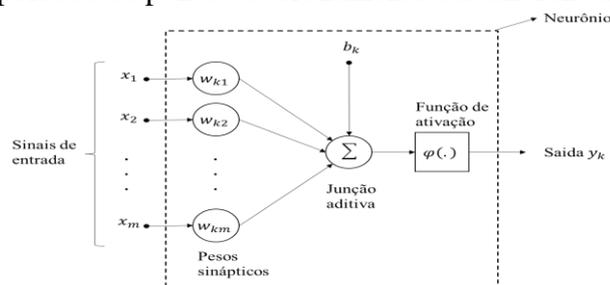

Figura 1 - Modelo de neurônio

Existem algumas variantes deste modelo para descrever o comportamento da unidade de processamento. Há também a possibilidade de variar a função de ativação conforme o comportamento desejado para a rede neural. Neste estudo o foco de análise são redes alimentadas diretamente com múltiplas camadas.

## 3.1 Redes Alimentadas Diretamente com Múltiplas Camadas

A presença de camadas ocultas é o principal fator diferencial deste tipo de arquitetura de rede representada na Figura 2. Os nós computacionais que se encontram nestas camadas são denominados neurônios ocultos. A presença de camadas ocultas dá à rede a capacidade de extrair estatísticas de ordem elevada, ou seja, com isso a rede neural captura correlações entre os *inputs* e as expressa nos *outputs*. De uma forma simples, a rede passa a englobar na resolução de um determinado problema todas as variáveis com seus respectivos pesos. De acordo com Hornik (1989), uma rede neural com algumas ou apenas uma camada e uma função de transferência sigmoide, é capaz de aproximar qualquer função mensurável de Borel de uma dimensão finita do espaço até outra com qualquer precisão desejável. Hagan, Demuth e Beale (2014) destacam que, em teoria, não há restrições para o número de camadas que uma rede neural pode ter, mas que, na prática, na maioria delas são criadas com três camadas: uma camada de entrada, uma de saída e apenas uma camada oculta ou escondida. A Figura 2 (a) representa uma rede alimentada diretamente com apenas uma camada oculta. Já a Figura 2 (b), representa uma rede alimentada diretamente com 2 camadas ocultas.

## 3.2 Otimização de Arquiteturas de Redes Neurais Artificiais

De acordo com Bernados (2006) existem quatro elementos que comprometem a arquitetura de uma rede neural artificial: (i) número de camadas; (ii) número de neurônios em cada camada; (iii) função de ativação de cada camada; (iv) algoritmo de treinamento. Infelizmente não há como determinar quantas camadas ocultas ou quantos neurônios ocultos precisamos para cada problema. As funções de ativação são muito dependentes dos tipos de dados disponíveis (binário, bipolar ou decimal).

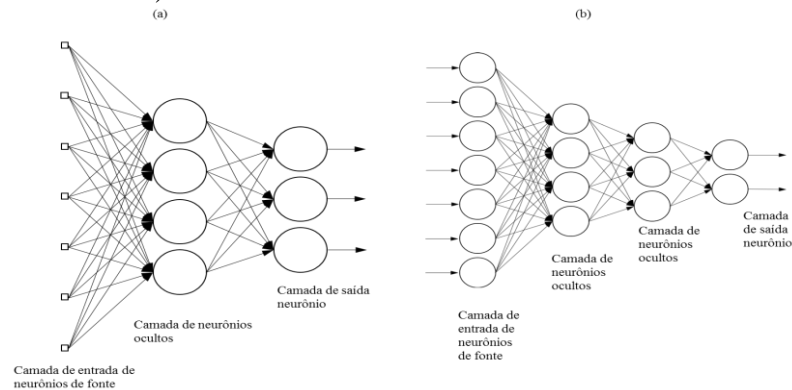

Figura 2 - Redes alimentadas diretamente com múltiplas camadas

Usualmente, utilizam-se funções sigmoides não (BENARDOS e VOSNIAKOS, 2006). Os algoritmos de treinamento influenciam basicamente a velocidade de treinamento de uma rede neural artificial ou o quanto de poder computacional é necessário. Bernados (2006) assume que os elementos mais críticos para a determinação de arquiteturas de redes acíclicas são o número de camadas e o número de neurônios, sendo tais características as principais responsáveis por afetar a capacidade da rede acíclica generalizar interpretação dos respectivos *inputs*. Com isso, a rede neural pode se tornar pouco capaz de aproximar uma correlação não linear com os dados de treinamento. O autor ainda destaca algumas formas de otimizar uma arquitetura.

## 4  Metodologia

Para determinar qual a melhor arquitetura de rede neural para o tipo de problema contemplado (previsão de retorno das ações), realizou-se uma série de testes variando-se parâmetros da arquitetura até que se chegasse a uma série de redes neurais que apresentassem melhores resultados. Métodos similares são utilizados em trabalhos como os de Kaastra (1996), no qual considerações sobre o número ótimo de camadas e o número ótimo de neurônios são abordadas. Aqui ampliamos o foco da metodologia de escolha de arquitetura, aprofundando mais nas características que a compõem e como impactam no desempenho de uma rede neural para o propósito específico de previsão dos retornos de ações.

### 4.1  Descrição do Problema

O objetivo deste estudo é determinar de que forma diferentes arquiteturas se comportam e como suas características impactam o desempenho das redes dentro do problema de previsão do retorno de ações negociadas na Bolsa de Valores de São Paulo (embora a metodologia possa ser estendida para o retorno de outros tipos de ativos). Neste trabalho é realizada uma busca exaustiva por uma arquitetura com desempenho assertivo superior. Para isso são criadas diversas arquiteturas de redes neurais através da alteração de parâmetros como: **número de camadas** e **número de neurônios**. Para cada arquitetura é realizado um treinamento através do método de retropropagação, utilizando as variáveis de mercado descritas a seguir como *input*, e a base histórica como referência para o aprendizado. Em seguida, são realizados testes comparando o resultado obtido pela previsão de retornos (*output* da rede neural) com o ocorrido de fato pela base histórica. São listadas as melhores arquiteturas do ponto de vista de assertividade, para que seja possível, visualizar padrões de acordo com as características das arquiteturas. Para validar algumas inferências, realizam-se testes estatísticos comparando a amostra das melhores arquiteturas de redes neurais com a população gerada pelo método exaustivo. Além disso, realizou-se uma regressão linear das características de arquitetura pelo erro cometido para que se estabelecesse alguma relação de causa e efeito entre uma arquitetura de rede neural e o seu desempenho. As variáveis de *input* (ou variáveis de entrada) são divididas em três grupos: (i) variáveis relacionadas diretamente ao retorno das ações, (ii) variáveis do setor econômico analisado e (iii) variáveis globais que podem ser associadas de diferentes formas com o retorno da ação. Destes *inputs*, esperamos como *output* (ou também saída) o retorno da ação que deveria resultar dos *inputs* daquele momento. O treinamento é feito pelo método de aprendizagem supervisionada utilizando os retornos reais históricos das ações escolhidas. A Figura 3 apresenta um modelo visual do problema analisado.

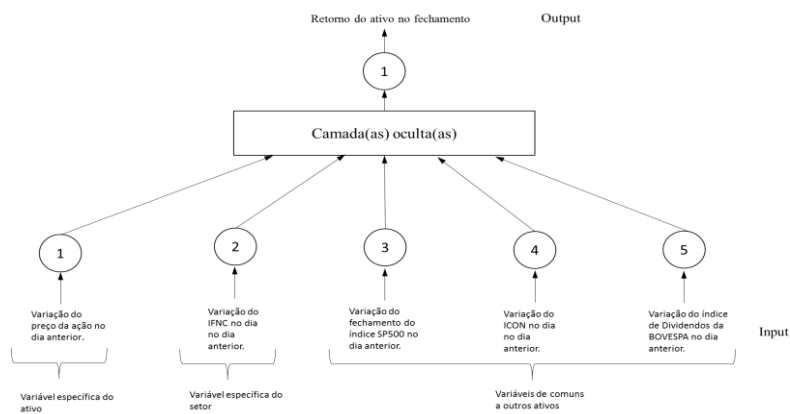

Figura 3 - Design do problema de redes neurais

## 4.2 Seleção de variáveis globais

Algumas variações de índices e retornos tiveram seleção prévia abrangendo os principais fatores econômicos conhecidos pelo mercado. As variáveis tomam como referência o dia que a ação será precificada, ou seja, se estamos prevendo o retorno de uma ação para certa data, utilizamos o retorno do dia anterior e a variação de índices escolhidos para que seja efetuada a previsão desejada. Sempre são utilizados como base os índices e preços no fechamento para fins de cálculo de variação do retorno. São as seguintes variáveis aqui consideradas:

1. **<u>RETORNO_ATIVO_FECHAMENTO</u>– Retorno do ativo no fechamento.** Variável dependente do problema analisado. Para uma certa data $t$ o retorno $R_{t-1,t}$, é dado por $\frac{P_t - P_{t-1}}{P_{t-1}}$. $P_t$ é o preço do ativo no fechamento no dia de referência. $P_{t-1}$ é o preço de fechamento do ativo na data anterior de referência. Esta é a variável que estamos tentando prever.
2. **<u>RETORNO_ATIVO_D-1</u> – Retorno da ação no dia anterior.** Trata-se do retorno do ativo no dia anterior à data de referência $R_{t-2,t-1}$. A forma como esta variável se relaciona a $R_{t-1,t}$, esta análise está mais associada às análises técnicas e não às análises fundamentalistas.
3. **<u>VARIACAO_INDICE_DE_DIVIDENDOS_D-1</u> – Variação do índice de dividendos da BOVESPA no dia anterior.** O cálculo do índice de dividendos é o resultado de uma carteira teórica, elaborada de acordo com os critérios estabelecidos pela metodologia da BM&FBOVESPA. Este índice utiliza procedimentos e regras constantes do manual de definições e procedimentos dos índices da BM&FBOVESPA. A utilização do índice de dividendos para previsão de retornos de ações é clássica, tendo como inspiração o artigo de Fama e French (1988).

4. **VARIACAO_SP500_D-1 – Variação do fechamento do índice SP500 no dia anterior.** A utilização desta variável buscar capturar efeitos de mercados externos no preço dos ativos brasileiros.
5. **VARIACAO_DO_ICON_D-1 - Variação do ICON no dia anterior.** ICON (Índice de consumo) é o resultado de uma carteira teórica de ativos, elaborada de acordo com os critérios estabelecidos na metodologia fornecida pela BM&FBOVESPA encontrados no Manual de Definições e Procedimentos dos Índices. A utilização desta variável busca capturar os impactos do consumo cíclico e não cíclico no retorno das ações. Este fator é utilizado no artigo de Lettay e Ludvigson (2001).
6. **VARIACAO_DO_IFNC_D-1 - Variação do IFNC no dia no dia anterior.** IFNC (Índice BM&FBOVESPA Financeiro) é o resultado de uma carteira teórica de ativos, elaborada de acordo com os critérios estabelecidos na metodologia fornecida pela BM&FBOVESPA encontrados no Manual de Definições e Procedimentos dos Índices. Utilizamos a variação do índice para mensurar parte do prêmio pelo risco do ativo analisado em relação ao mercado financeiro (FAMA e FRENCH, 1989).

Para treinamento da rede foram utilizados os dados históricos dos anos 2013 e 2014. Foram utilizados os dados históricos do ano de 2015 para a validação dos resultados e cálculo do erro cometido pela rede. Todos os dados foram extraídos da base de dados da Economática e são disponibilizados diariamente para os dias úteis da bolsa de São Paulo. Aqui usamos para cálculo dos retornos e da variação o fechamento do preço dos ativos e o fechamento dos índices analisados.

### 4.3 Seleção e Construção de Arquiteturas de Rede Neural

O algoritmo utilizado para geração de arquiteturas de redes neurais é o método de *busca por força bruta,* também conhecido por *busca exaustiva*. Embora seja essencialmente utilizado para problemas com complexidade menor, é capaz de gerar um volume adequado para a análise da eficiência de diferentes arquiteturas de forma mais eficiente. O número de redes testado é determinado pelo número máximo de neurônios e o número máximo de camadas adotado. Para determinado número de camadas $K$, foram testadas todas as redes possíveis para um número máximo de neurônios $N$ em cada camada. Como neste trabalho testamos diferentes números de camadas, a variável número de camadas precisa ser alterada até seu valor máximo adotado. Portanto, para calcular o número total de redes testadas precisamos realizar a somatória do número máximo de neurônios por camada elevado ao número de camadas considerado variando-se o número de camadas até o seu valor máximo adotado, $T$, dado por $T = \sum_{i=1}^{K} N^i$, em que, $K$ é o número de camadas máximo, e $N$ o número máximo de neurônios por camada. O critério para escolha das arquiteturas foi a execução exaustiva de todas as arquiteturas que contemplassem no máximo 6 neurônios por camada ($N = 6$) e no máximo 5 camadas ($K = 5$), portanto foram testadas 9.330 arquiteturas diferentes. Estes números foram adotados para que em termos de processamento e memória computacionais fossem executáveis dado o poder de processamento disponível.

### 4.4 Critérios de Avaliação das Redes

Para que possamos classificar as redes e realizar comparações em termos de suas arquiteturas, utilizamos as seguintes características das arquiteturas de redes neurais para avaliação do seu desempenho assertivo: (i)Número de camadas (N_CAMADAS); (ii) Números de neurônios (N_NEURONIOS); (iii) Média do número de neurônios por camada (N_MED_NEURONIOS); (iv) Desvio padrão do número de neurônios por camada em relação ao número médio de neurônios na arquitetura considerada (DESVIO_N_NEURONIOS); (v) Número de mudanças de sentido de crescimento do número de neurônios entre camadas. Utilizaremos o termo "número de inflexões" para se referir a esta característica (N_INFLEXOES). Vale ressaltar que o número de inflexões é o parâmetro menos usual apresentado nos trabalhos estudados. Esta característica reflete um grau de uniformidade da rede em relação ao aumento ou decréscimo do número de neurônios de uma camada para outra. Abaixo, a Figura 4 ilustra como exemplo este conceito.

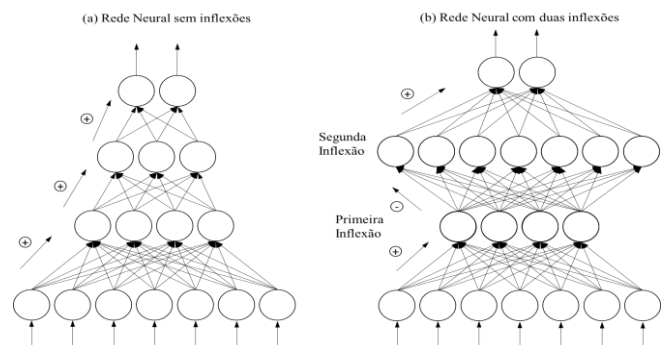

Figura 4 - Exemplo de redes com e sem inflexões

A Figura 4 (a) não apresenta nenhuma inflexão, pois de uma camada para outra a variação do número de neurônios não muda de sinal, ou seja, há sempre redução do número de neurônios. Caso houvesse apenas aumento do número de neurônios a cada camada, também não haveria inflexões. Já na Figura 4 (b) observa-se que há mudanças de uma camada para outra do sinal da variação do número de neurônios, significando que em um momento observa-se um decrescimento deste número e em seguida um acréscimo, e vice-versa.

### 4.5 Testes estatísticos

Foram realizados testes estatísticos utilizando a distribuição normal e a distribuição binominal para comparar as amostras selecionadas com as arquiteturas que tiveram o melhor desempenho, e comparou-se esta amostra com a população para que fosse possível identificar qualquer característica que diferenciasse a amostra da população. Nomeou-se neste trabalho como ***top 40 arquiteturas*** a amostra das 40 melhores arquiteturas. Utilizou-se com critério para determinar o tamanho da amostra de 40, as arquiteturas de rede que tiveram um erro de até 2% em relação a que teve o melhor desempenho e que fosse de fácil exibição no trabalho

### 4.6 Regressão linear das características de arquitetura

A partir das características consideradas, iremos observar se o desempenho das diferentes redes se correlaciona com as características das redes, e de qual forma. Para tal análise, foi utilizado uma Regressão Linear Múltipla (RLM). Embora a correlações aparentemente não seja linear, esta é uma aproximação para identificar quais são os fatores que impactam de forma significativa a assertividade das redes neurais. Para que evitar problemas de multicolinearidade, apresentamos a matriz de correlação entre as variáveis explicativas do problema. Excluindo as variáveis que possuem grande correlação com outras variáveis, a regressão linear é realizada. Iremos também executar regressões lineares utilizando variáveis explicativas elevadas ao quadrado com o intuito de capturar qualquer relação não linear destas variáveis com o erro da respectiva regressão.

## 5 Resultados

Apresentamos a seguir uma série de resultados provenientes dos experimentos realizados a partir de testes empíricos, aplicando a metodologia proposta no capítulo anterior. Para as variáveis selecionadas foram testadas diversas redes neurais que tiveram seus parâmetros de arquitetura modificados variando-se o número de neurônios e o número de camadas. Para este fim, foram selecionados o seguinte ativo: BBAS3 (ação preferencial do Banco do Brasil SA negociada na BM&FBOVESPA). Um estudo comparativo, aplicando a mesma metodologia aqui proposta, é apresentado em Felizardo (2017). Quando nos referirmos às melhores arquiteturas iremos utilizar o termo ***top 40* arquiteturas**. Quando nos referirmos à população de todas as arquiteturas de redes geradas a partir de um determinado conjunto de parâmetros, no nosso caso, as 9.330 arquiteturas, utilizaremos o termo **população**. A representação de cada arquitetura será feita através de um *label* da forma "$n_1.n_2.n_3...n_n$", em que $1 \leq n_i \leq N$, $i = 1,...,K$, com $n_i$ indicando o número de neurônios na camada $i$, e $k$ o número de camadas da arquitetura.

### 5.1 Previsão do Retorno do ativo

A seguir são apresentadas as melhores arquiteturas em termos de assertividade usando o ativo BBAS3. Os resultados são mostrados de diversas formas de visualização para se identificar padrões relevantes de desempenho.

### 5.2 Ranking das melhores arquiteturas

Com o objetivo de facilitar o reconhecimento de padrões presentes nas diversas arquiteturas propostas, aquelas que se mostraram superiores na assertividade, foi criada a na Tabela 1 com as arquiteturas que produziram os quarenta melhores desempenhos para suas respectivas ações. Na Tabela 1 é apresentado o desempenho relativo das redes neurais comparadas com aquela que obteve o melhor desempenho, usada como referência. A coluna "Assertividade Relativa (%)" apresenta a piora relativa da assertividade das arquiteturas em relação à melhor arquitetura de rede. Adota-se o erro absoluto de 0,2083% como valor de referência dado que é o menor erro obtido pelas diferentes arquiteturas testadas para o ativo BBAS3, no caso, para a arquitetura"1.3.6.3.1". O segundo menor erro absoluto foi de 0,2086% para arquitetura "4.4.1.5.2". A assertividade relativa desta arquitetura é obtida a partir da divisão do seu erro absoluto pelo menor erro absoluto subtraída de uma unidade, obtendo-se uma diferença relativa entre eles de 0,0571%. São comparadas todas as arquiteturas com a arquitetura de melhor desempenho para obtermos as diferenças relativas ao seu erro absoluto. Percebeu-se um erro relativo de no máximo 46% entre a arquitetura que teve o melhor desempenho e a arquitetura que teve o pior desempenho em toda a população. Também foi identificado que há uma diferença relativa média da ordem de 5,5%, e com desvio padrão de 3,5%, o que revela que há pouca diferença relativa entre o desempenho das redes. Na Tabela 3 ordenamos de forma lexicográfica (ordenando de forma crescente pelo número de neurônio de cada camada) as *top 40* arquiteturas de tal forma que facilite a percepção de padrões de arquiteturas de rede. É difícil identificar qualquer relação entre o número de neurônios na primeira camada e o desempenho da rede neural, não conseguindo extrair qualquer observação relevante nesta ordenação. Na Tabela 4 é possível visualizar como é a distribuição das arquiteturas em relação ao número de camadas.

Tabela 1 – *Top 40* arquiteturas na ordem de desempenho – BBSA3

| Posição no Ranking | Arquiteturas | Erro Absoluto (%) | Assertividade Relativa (%) | Posição no Ranking | Arquiteturas | Erro Absoluto (%) | Assertividade Relativa (%) |
|---|---|---|---|---|---|---|---|
| 1 | 1.3.6.3.1 |  | 0 | 21 | 3.4.1.1.2 | 0,2071 | 0,7378 |
| 2 | 4.4.1.5.2 | 0,2057 | 0,0571 | 22 | 4.1.1.4.3 | 0,2071 | 0,7467 |
| 3 | 6.3.4.4.4 | 0,206 | 0,1997 | 23 | 1.5.4.6.4 | 0,2072 | 0,762 |
| 4 | 5.6.6.3.5 | 0,2061 | 0,2561 | 24 | 1.1.5.4.6 | 0,2072 | 0,7736 |
| 5 | 6.2.4.1.1 | 0,2062 | 0,2922 | 25 | 3.3.4.6 | 0,2072 | 0,7738 |
| 6 | 5.3.6.3 | 0,2065 | 0,4194 | 26 | 4.2.4.5.6 | 0,2073 | 0,8232 |
| 7 | 6.4.4.1.6 | 0,2065 | 0,4509 | 27 | 3.5.5.5.5 | 0,2073 | 0,8408 |
| 8 | 5.2.4.6.4 | 0,2066 | 0,4723 | 28 | 1.1.3.6.6 | 0,2073 | 0,8515 |
| 9 | 3.5.5.4.1 | 0,2066 | 0,4751 | 29 | 6.2.1.5.1 | 0,2074 | 0,8634 |
| 10 | 5.2.1.1.2 | 0,2066 | 0,507 | 30 | 1.2.6.2.4 | 0,2074 | 0,8655 |
| 11 | 2.6.2.2.4 | 0,2067 | 0,5285 | 31 | 6.5.4.1.3 | 0,2074 | 0,8677 |
| 12 | 1.1.2.3.5 | 0,2067 | 0,5514 | 32 | 3.6.5.2.5 | 0,2074 | 0,8688 |
| 13 | 1.5.6.5.4 | 0,2068 | 0,5888 | 33 | 6.4.2.1.3 | 0,2074 | 0,873 |
| 14 | 1.4.2.3.5 | 0,2068 | 0,5892 | 34 | 1.6.3.2.3 | 0,2074 | 0,8748 |
| 15 | 2.5.5.6.6 | 0,2068 | 0,5898 | 35 | 2.5.2.4.5 | 0,2074 | 0,8961 |
| 16 | 5.4.5.6.3 | 0,2068 | 0,6021 | 36 | 1.3.6.5.5 | 0,2075 | 0,9059 |
| 17 | 5.6.3.3.6 | 0,2069 | 0,6255 | 37 | 1.3.2.5 | 0,2075 | 0,9094 |
| 18 | 6.4.6.4.6 | 0,207 | 0,6816 | 38 | 3.5.1.3.6 | 0,2075 | 0,9129 |
| 19 | 5.5.2.4.4 | 0,2071 | 0,7316 | 39 | 5.6.2.3.3 | 0,2075 | 0,9164 |
| 20 | 4.4.2.1.6 | 0,2071 | 0,7332 | 40 | 6.3.4.3.3 | 0,2075 | 0,9187 |

Tabela 2 - *Top 40* arquiteturas ordenadas de acordo com o número de inflexões

| Posição no Ranking | Arquiteturas | Erro Absoluto (%) | Número de inflexões | Posição no Ranking | Arquiteturas | Erro Absoluto (%) | Assertividade Relativa (%) |
|---|---|---|---|---|---|---|---|
| 9 | 3.5.5.4.1 | 0,2103 | 0 | 33 | 6.4.2.1.3 | 0,212 | 1 |
| 10 | 5.2.1.1.2 | 0,2104 | 0 | 36 | 1.3.6.5.5 | 0,2121 | 1 |
| 12 | 1.1.2.3.5 | 0,2106 | 0 | 2 | 4.4.1.5.2 | 0,2086 | 2 |
| 15 | 2.5.5.6.6 | 0,2108 | 0 | 5 | 6.2.4.1.1 | 0,2095 | 2 |
| 25 | 3.3.4.6 | 0,2115 | 0 | 6 | 5.3.6.3 | 0,2101 | 2 |
| 27 | 3.5.5.5.5 | 0,2118 | 0 | 8 | 5.2.4.6.4 | 0,2103 | 2 |
| 28 | 1.1.3.6.6 | 0,2119 | 0 | 14 | 1.4.2.3.5 | 0,2108 | 2 |
| 1 | 1.3.6.3.1 | 0,2083 | 1 | 16 | 5.4.5.6.3 | 0,2108 | 2 |
| 3 | 6.3.4.4.4 | 0,2091 | 1 | 24 | 1.1.5.4.6 | 0,2115 | 2 |
| 4 | 5.6.6.3.5 | 0,2094 | 1 | 29 | 6.2.1.5.1 | 0,2119 | 2 |
| 7 | 6.4.4.1.6 | 0,2102 | 1 | 30 | 1.2.6.2.4 | 0,2119 | 2 |
| 11 | 2.6.2.2.4 | 0,2105 | 1 | 32 | 3.6.5.2.5 | 0,2119 | 2 |
| 13 | 1.5.6.5.4 | 0,2108 | 1 | 34 | 1.6.3.2.3 | 0,212 | 2 |
| 17 | 5.6.3.3.6 | 0,2109 | 1 | 35 | 2.5.2.4.5 | 0,2121 | 2 |
| 19 | 5.5.2.4.4 | 0,2114 | 1 | 37 | 1.3.2.5 | 0,2121 | 2 |
| 20 | 4.4.2.1.6 | 0,2114 | 1 | 38 | 3.5.1.3.6 | 0,2121 | 2 |
| 21 | 3.4.1.1.2 | 0,2114 | 1 | 39 | 5.6.2.3.3 | 0,2121 | 2 |
| 22 | 4.1.1.4.3 | 0,2114 | 1 | 40 | 6.3.4.3.3 | 0,2122 | 2 |
| 26 | 4.2.4.5.6 | 0,2118 | 1 | 18 | 6.4.6.4.6 | 0,2112 | 3 |
| 31 | 6.5.4.1.3 | 0,2119 | 1 | 23 | 1.5.4.6.4 | 0,2115 | 3 |

Tabela 3 – *Top 40* arquiteturas ordenadas e agrupadas por ordem lexicográfica – BBSA3

| Posição no Ranking | Arquiteturas | Erro Absoluto (%) | Assertividade Relativa (%) | Posição no Ranking | Arquiteturas | Erro Absoluto (%) | Assertividade Relativa (%) |
|---|---|---|---|---|---|---|---|
| 12 | 1.1.2.3.5 | 0,2067 | 0,5514 | 22 | 4.1.1.4.3 | 0,2071 | 0,7467 |
| 28 | 1.1.3.6.6 | 0,2073 | 0,8515 | 26 | 4.2.4.5.6 | 0,2073 | 0,8232 |
| 24 | 1.1.5.4.6 | 0,2072 | 0,7736 | 2 | 4.4.1.5.2 | 0,2057 | 0,0571 |
| 30 | 1.2.6.2.4 | 0,2074 | 0,8655 | 20 | 4.4.2.1.6 | 0,2071 | 0,7332 |
| 37 | 1.3.2.5 | 0,2075 | 0,9094 | 10 | 5.2.1.1.2 | 0,2066 | 0,507 |
| 1 | 1.3.6.3.1 | - | 0 | 8 | 5.2.4.6.4 | 0,2066 | 0,4723 |
| 36 | 1.3.6.5.5 | 0,2075 | 0,9059 | 6 | 5.3.6.3 | 0,2065 | 0,4194 |
| 14 | 1.4.2.3.5 | 0,2068 | 0,5892 | 16 | 5.4.5.6.3 | 0,2068 | 0,6021 |
| 23 | 1.5.4.6.4 | 0,2072 | 0,762 | 19 | 5.5.2.4.4 | 0,2071 | 0,7316 |
| 13 | 1.5.6.5.4 | 0,2068 | 0,5888 | 39 | 5.6.2.3.3 | 0,2075 | 0,9164 |
| 34 | 1.6.3.2.3 | 0,2074 | 0,8748 | 17 | 5.6.3.3.6 | 0,2069 | 0,6255 |
| 35 | 2.5.2.4.5 | 0,2074 | 0,8961 | 4 | 5.6.6.3.5 | 0,2061 | 0,2561 |
| 15 | 2.5.5.6.6 | 0,2068 | 0,5898 | 29 | 6.2.1.5.1 | 0,2074 | 0,8634 |
| 11 | 2.6.2.2.4 | 0,2067 | 0,5285 | 5 | 6.2.4.1.1 | 0,2062 | 0,2922 |
| 25 | 3.3.4.6 | 0,2072 | 0,7738 | 40 | 6.3.4.3.3 | 0,2075 | 0,9187 |
| 21 | 3.4.1.1.2 | 0,2071 | 0,7378 | 3 | 6.3.4.4.4 | 0,206 | 0,1997 |
| 38 | 3.5.1.3.6 | 0,2075 | 0,9129 | 33 | 6.4.2.1.3 | 0,2074 | 0,873 |
| 9 | 3.5.5.4.1 | 0,2066 | 0,4751 | 7 | 6.4.4.1.6 | 0,2065 | 0,4509 |
| 27 | 3.5.5.5.5 | 0,2073 | 0,8408 | 18 | 6.4.6.4.6 | 0,207 | 0,6816 |
| 32 | 3.6.5.2.5 | 0,2074 | 0,8688 | 31 | 6.5.4.1.3 | 0,2074 | 0,8677 |

Tabela 4 - *Top 40* arquiteturas ordenadas pelo número de camadas

| Posição no Ranking | Arquiteturas | Erro Absoluto (%) | Número de Camadas | Posição no Ranking | Arquiteturas | Erro Absoluto (%) | Número de Camadas |
|---|---|---|---|---|---|---|---|
| 6  | 5 . 3 . 6 . 3     | 0,2101 | 4 | 19 | 5 . 5 . 2 . 4 . 4 | 0,2114 | 5 |
| 25 | 3 . 3 . 4 . 6     | 0,2115 | 4 | 20 | 4 . 4 . 2 . 1 . 6 | 0,2114 | 5 |
| 37 | 1 . 3 . 2 . 5     | 0,2121 | 4 | 21 | 3 . 4 . 1 . 1 . 2 | 0,2114 | 5 |
| 1  | 1 . 3 . 6 . 3 . 1 | 0,2083 | 5 | 22 | 4 . 1 . 1 . 4 . 3 | 0,2114 | 5 |
| 2  | 4 . 4 . 1 . 5 . 2 | 0,2086 | 5 | 23 | 1 . 5 . 4 . 6 . 4 | 0,2115 | 5 |
| 3  | 6 . 3 . 4 . 4 . 4 | 0,2091 | 5 | 24 | 1 . 1 . 5 . 4 . 6 | 0,2115 | 5 |
| 4  | 5 . 6 . 6 . 3 . 5 | 0,2094 | 5 | 26 | 4 . 2 . 4 . 5 . 6 | 0,2118 | 5 |
| 5  | 6 . 2 . 4 . 1 . 1 | 0,2095 | 5 | 27 | 3 . 5 . 5 . 5 . 5 | 0,2118 | 5 |
| 7  | 6 . 4 . 4 . 1 . 6 | 0,2102 | 5 | 28 | 1 . 1 . 3 . 6 . 6 | 0,2119 | 5 |
| 8  | 5 . 2 . 4 . 6 . 4 | 0,2103 | 5 | 29 | 6 . 2 . 1 . 5 . 1 | 0,2119 | 5 |
| 9  | 3 . 5 . 5 . 4 . 1 | 0,2103 | 5 | 30 | 1 . 2 . 6 . 2 . 4 | 0,2119 | 5 |
| 10 | 5 . 2 . 1 . 1 . 2 | 0,2104 | 5 | 31 | 6 . 5 . 1 . 1 . 3 | 0,2119 | 5 |
| 11 | 2 . 6 . 2 . 2 . 4 | 0,2105 | 5 | 32 | 3 . 6 . 5 . 2 . 5 | 0,2119 | 5 |
| 12 | 1 . 1 . 2 . 3 . 5 | 0,2106 | 5 | 33 | 6 . 4 . 2 . 1 . 3 | 0,212  | 5 |
| 13 | 1 . 5 . 6 . 5 . 4 | 0,2108 | 5 | 34 | 1 . 6 . 3 . 2 . 3 | 0,212  | 5 |
| 14 | 1 . 4 . 2 . 3 . 5 | 0,2108 | 5 | 35 | 2 . 5 . 2 . 4 . 5 | 0,2121 | 5 |
| 15 | 2 . 5 . 5 . 6 . 6 | 0,2108 | 5 | 36 | 1 . 3 . 6 . 5 . 5 | 0,2121 | 5 |
| 16 | 5 . 4 . 5 . 6 . 3 | 0,2108 | 5 | 38 | 3 . 5 . 1 . 3 . 6 | 0,2121 | 5 |
| 17 | 5 . 6 . 3 . 3 . 6 | 0,2109 | 5 | 39 | 5 . 6 . 2 . 3 . 3 | 0,2121 | 5 |
| 18 | 6 . 4 . 6 . 4 . 6 | 0,2112 | 5 | 40 | 6 . 3 . 4 . 3 . 3 | 0,2122 | 5 |

Observa-se que arquiteturas que possuem um número mais elevado de camadas aparentemente se encontram mais presentes entre as melhores redes neurais. Para que esta hipótese seja plausível, é necessário a aplicação de um teste estatístico que é apresentado mais à frente. Já na Tabela 5, ordenamos as redes neurais pelo número de inflexões tal que seja possível identificar algum padrão relativo ao número de inflexões mais prevalentes nas arquiteturas melhor classificadas. É possível notar que, redes que possuem três inflexões ou mais são menos prevalentes. Isso pode ser devido à população testada. Portanto, para confirmar se as *top 40* arquiteturas de redes neurais se diferenciam de alguma forma da população, é necessário realizar um teste de hipótese, também apresentado a seguir no item 0.

### 5.3 Testes Estatísticos

No que segue, são apresentados testes estatísticos que têm como objetivo evidenciar se de fato as *top 40* arquiteturas possuem algumas características que se diferenciam da população total das arquiteturas testadas.

**Número de Camadas**

Como foi definido anteriormente, no item 4, o número máximo de camadas ocultas adotado foi 5. Com isso, para um dado conjunto de parâmetros, podemos calcular a proporção $p$ das arquiteturas encontradas na população e a proporção $p_0$, nas arquiteturas testadas na **top 40 arquiteturas**, testando se $p_0 = p$. Na Tabela 5 temos a probabilidade associada às arquiteturas com um determinado número de camadas dentro de uma população. Para se obter a probabilidade de se encontrar dentro da amostra **top 40 arquiteturas** com um determinado número de camadas, precisamos obter a proporção de arquiteturas com este número de camadas dentro da amostra em relação à proporção do número de arquiteturas com este número de camadas na população. Tomemos como exemplo o cálculo da probabilidade de ocorrência de arquiteturas nas top 40 que possuam 4 camadas ocultas (vide linha destacada na Tabela 5). A proporção do número de arquiteturas com este número de camadas ocultas foi de 0,0750 (dada por 3/40, pois se observaram 3 arquiteturas com 4 camadas ocultas nas top 40 para este ativo). O número de arquiteturas com no máximo 6 neurônios por camada e 4 camadas ocultas é dado por $6^4$=1.296. Logo, a proporção de arquiteturas com este número de camadas na população total, $p$, é de 0,1389 (dada por 1.296/9.330). Calculando-se a probabilidade, $f(k; n, p) =$ $\binom{n}{k} p^k (1-p)^{n-k}$, de se encontrar numa amostra de tamanho *n*, no caso igual a 40, um número de arquiteturas *k*, no caso igual a 3 (número de arquiteturas com 4 camadas ocultas encontradas nas top 40) de 0,1047 ou 10,47%. Adotando um nível de significância de 10%, não podemos rejeitar o fato de que a distribuição da amostra é igual à distribuição da população para as características e ativos analisados. Observando a proporção da amostra, a proporção da população e a probabilidade exibidas na Tabela 5, verifica-se que para um maior número de camadas a proporção da amostra apresenta-se inferior à proporção da população, possuindo baixa probabilidade de que sejam estatisticamente iguais. Aparentemente, há um decréscimo do desempenho assertivo com o aumento do número de camadas. Portanto, tal observação pode ser um indicativo de que quanto mais camadas adicionarmos à rede neural, pior será seu desempenho assertivo. Aparentemente, há um decréscimo do desempenho assertivo com o aumento do número de camadas.

Tabela 5 – Probabilidade associado ao número de camadas da população.

| Número de Camadas | Número de Arquiteturas na Amostra | Proporção da Amostra | Proporção da População | Probabilidade |
|---|---|---|---|---|
| 1 | 0 | 0,0000 | 0,0006 | 0,9746 |
| 2 | 0 | 0,0000 | 0,0039 | 0,8567 |
| 3 | 0 | 0,0000 | 0,0232 | 0,3918 |
| 4 | 3 | 0,0750 | 0,1389 | 0,1047 |
| 5 | 33 | 0,8250 | 0,8334 | 0,1623 |

**Número médio de neurônios por camada**

Com o intuito de comparar as *Top 40* arquiteturas com a população, também é realizada uma análise assumindo-se uma distribuição normal para o número médio de neurônios por camada, dado que esse valor se trata da somatória das médias do número de neurônios em cada camada. A Tabela 6 representa, de forma análoga ao já apresentado anteriormente, a probabilidade de obtermos uma amostra com determinado número médio de neurônios, revelando se de fato as *top 40* arquiteturas têm um desempenho superior à da população devido ao número médio de neurônios. Aparentemente, mesmo considerando uma significância de 10%, não é provável que esta amostra tenha se diferenciado da população por este aspecto, ou seja rejeitamos a hipótese nula da média da amostra ser igual à média da população. Portanto, não há indícios de que o a média do número de neurônios por camada dentro de uma arquitetura, seja de fato um parâmetro relevante para explicar o desempenho assertivo da rede neural.

Tabela 6 - Distribuição Normal – média de neurônios das Top 40 arquiteturas.

| Média da Amostra | Desvio padrão da População | p-valor |
|---|---|---|
| 1,4991 | 0,4264 | 0,50 |

**Desvio padrão do número de neurônios**

De forma análoga ao que foi feito na anteriormente, a Tabela 7**Erro! Fonte de referência não encontrada.** apresenta a probabilidade calculada utilizando-se a distribuição normal. Neste caso a característica analisada é o desvio padrão do número de neurônios.

Novamente rejeitamos a hipótese nula de que a média da amostra é igual à média da população, sendo assim possível perceber que também para este caso não há diferença entre a média da população e as *top 40* arquiteturas. Temos o indicativo de que o desvio padrão do número de neurônios não é relevante para explicar o erro cometido por uma rede neural.

Tabela 7 - Desvio padrão do número de neurônios médio comparado à população.

| Média da Amostra | Desvio padrão da População | p-valor |
|---|---|---|
| 3,65 | 0,7858 | 0,50 |

**Número de inflexões**

Na Tabela 8 é apresentado o número de redes com um determinado número de inflexões que se encontram entre aquelas redes neurais que tiveram melhor desempenho assertivo. Novamente, as primeiras colunas apresentam quantas redes neurais com 0, 1, 2 e 3 inflexões estão na seleção das *Top 40* arquiteturas. Calculamos a probabilidade de a proporção da amostra ser igual a proporção da população, utilizando-se da equação da distribuição binomial previamente apresentada. Com 10% de significância, os únicos números de inflexões que apresentaram certa discrepância em relação à população foram os de 2 e 3 inflexões. Neste caso, ainda é possível perceber que ao possuir 3 inflexões a rede tem desempenho pior. Porém, de forma contrária ao esperado, para 2 inflexões as redes aparentemente apresentam um desempenho superior. Seria de se esperar que quanto mais inflexões a arquitetura de rede neural possui, pior será o seu desempenho, dado que é uma medida de uniformidade da rede neural. Para o número de inflexões máximo analisado, não é possível fazer qualquer inferência com relação ao comportamento do desempenho assertivo em função do número de inflexões dado que temos um comportamento não linear. É possível que para o número de inflexões analisado, não seja possível de fato verificar qualquer relação entre o número de inflexões e o erro cometido por uma determinada arquitetura

Tabela 8 - Teste da amostra Top 40 arquiteturas do número de inflexões.

| Número de Inflexões | Número de Arquiteturas na Amostra | Proporção da Amostra | Proporção da População | Probabilidade |
|---|---|---|---|---|
| 0 | 7 | 0,175 | 0,1792 | 0,1636 |
| 1 | 14 | 0,35 | 0,3588 | 0,1304 |
| 2 | 15 | 0,375 | 0,3181 | 0,0968 |
| 3 | 2 | 0,05 | 0,1438 | 0,0442 |

**Número de neurônios total**

Dentre as *Top 40* arquiteturas a Tabela 9 mostra o número de ocorrências de arquiteturas com um determinado número de neurônios. Na mesma tabela apresenta-se a média de ocorrência de uma determinada arquitetura com um determinado número de neurônios e sob esta média é calculada a probabilidade que compara duas proporções, a dos *Top 40* arquiteturas (amostra) e a população. Esta análise é semelhante à realizada para o número de camadas. Apresentamos ao lado da tabela um histograma para facilitar a visualização desta inferência. Esta observação pode ser um indicio de que haja de fato um intervalo de número de neurônios em que o desempenho assertivo das redes neurais é pior.

### 5.4 Regressão Linear Múltpla

Com o intuito de verificar se há alguma correlação (linear) entre os parâmetros de arquitetura considerados e o desempenho assertivo das redes, é apresentada uma regressão linear múltpla. Esta primeira análise tem como objetivo tentar identificar qualquer forma de relação entre as características e a sua assertividade. Antes de realizar a regressão linear, apresentamos a matriz de correlações das variáveis explicativas (definidas no item 4.6) e a variável ERRO DA PREVISAO que se trata da somatória dos erros quadráticos da previsão (variável dependente) para que evitemos problemas de multicolinearidade. É possível perceber pela matriz que, como esperado, o número de neurônios tem grande correlação com o número médio de neurônios e, portanto, na regressão linear, excluiremos o número médio de neurônios. Na Tabela 11 as variáveis explicativas são as apresentadas na primeira coluna (e novamente listadas abaixo), a variável dependente é a soma dos erros quadráticos entre o retorno do ativo previsto pela rede e o correspondente retorno do ativo observado. Os coeficientes da RLM tentam revelar como algumas características influenciam o desempenho das redes neurais. O desvio padrão apresentado na terceira coluna é uma medida de dispersão desta variável explicativa, podendo ser usado para entender o quão abrangente é a população. Na quinta coluna, é apresentado o *p-valor*, que é a probabilidade para que coeficiente apresentado seja igual a zero dado um intervalo com 95% de confiança, apresentado na última coluna. Caso o coeficiente não se mostre diferente de zero pelo p-valor, significa que a variável explicativa em análise não possui uma correlação linear com a variável dependente.

Tabela 9 - Probabilidade de ocorrência de um determinado número de neurônios nas *Top 40* arquiteturas – BBSA3.

| Número de Neurônios | Número de Arquiteturas na Amostra | Proporção da Amostra | Proporção da População | Probabilidade |
|---|---|---|---|---|
| 1 | 0 | 0 | 0,0001 | 0,9957 |
| 2 | 0 | 0 | 0,0002 | 0,9915 |
| 3 | 0 | 0 | 0,0004 | 0,9830 |
| 4 | 0 | 0 | 0,0009 | 0,9663 |
| 5 | 0 | 0 | 0,0017 | 0,9336 |
| 6 | 0 | 0 | 0,0033 | 0,8754 |
| 7 | 0 | 0 | 0,0060 | 0,7860 |
| 8 | 0 | 0 | 0,0103 | 0,6612 |
| 9 | 0 | 0 | 0,0166 | 0,5117 |
| 10 | 0 | 0 | 0,0253 | 0,3589 |
| 11 | 2 | 0,05 | 0,0362 | 0,2519 |
| 12 | 1 | 0,025 | 0,0489 | 0,2770 |
| 13 | 1 | 0,025 | 0,0623 | 0,2029 |
| 14 | 0 | 0 | 0,0751 | 0,0440 |
| 15 | 3 | 0,075 | 0,0859 | 0,2257 |
| 16 | 2 | 0,05 | 0,0928 | 0,1659 |
| 17 | 3 | 0,075 | 0,0951 | 0,2107 |
| 18 | 2 | 0,05 | 0,0923 | 0,1677 |
| 19 | 3 | 0,075 | 0,0848 | 0,2270 |
| 20 | 2 | 0,05 | 0,0735 | 0,2315 |
| 21 | 2 | 0,05 | 0,0600 | 0,2674 |
| 22 | 0 | 0 | 0,0461 | 0,1515 |
| 23 | 1 | 0,025 | 0,0331 | 0,3562 |
| 24 | 0 | 0 | 0,0221 | 0,4094 |
| 25 | 0 | 0 | 0,0135 | 0,5805 |
| 26 | 0 | 0 | 0,0075 | 0,7399 |
| 27 | 0 | 0 | 0,0038 | 0,8604 |
| 28 | 0 | 0 | 0,0016 | 0,9377 |
| 29 | 0 | 0 | 0,0005 | 0,9788 |
| 30 | 0 | 0 | 0,0001 | 0,9957 |

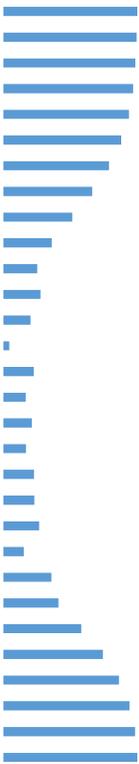

Tabela 10 - Matriz de correlação das características da arquitetura

| | ERRO | N_CAMADAS | N_NEURONIOS | N_MED_NEURONIOS | DESVIO_N_NEURONIOS | N_INFLEXOES |
|---|---|---|---|---|---|---|
| ERRO DA PREVISÃO | 1,0000 | | | | | |
| N_CAMADAS | -0,0553 | 1,0000 | | | | |
| N_NEURONIOS | 0,1172 | 0,4143 | 1,0000 | | | |
| N_MED_NEURONIOS | 0,1566 | 0,0000 | 0,9028 | 1,0000 | | |
| DESVIO_N_NEURONIOS | -0,0260 | 0,1222 | 0,0506 | 0,0000 | 1,0000 | |
| N_INFLEXOES | -0,0212 | 0,2599 | 0,1076 | 0,0000 | 0,1387 | 1,0000 |

É possível perceber que o número de inflexões aparentemente não há uma correlação linear com a somatória do erro quadrático das redes neurais. Já o número de camadas, revela pelo coeficiente um comportamento diferente do observado na amostra, ou seja, pelo que é observado, quanto maior o número de camadas, melhor será o desempenho assertivo da rede neural.

Tabela 11 - Regressão linear do parâmetro de arquitetura pelo desempenho da rede neural

| Variável explicativa | Coeficiente | Desvio Padrão | t | p-valor>|t| | Intervalo com 95% de confiança |
|---|---|---|---|---|---|
| n_camadas | -0,0052 | 0,0005 | -10,50 | 0,0000 | [-0,0062003, -0,0042494] |
| n_neuronios | 0,0009 | 0,0001 | 15,07 | 0,0000 | [0,0007476, 0,0009711] |
| desvio_n_neuronios | -0,0009 | 0,0005 | -1,83 | 0,0670 | [-0,0019245, 0,0000643] |
| n_inflexoes | -0,0001 | 0,0002 | -0,49 | 0,6260 | [-0,0005764, 0,0003471] |
| _cons | 0,2471 | 0,0022 | 112,97 | 0,0000 | [0,2427735, 0,2513472] |

O número de neurônios também teve comportamento, não esperado, pois esperava-se que quanto mais neurônios menor fosse o erro, que é o oposto ao resultado apresentado. Quando analisamos o desvio padrão do número de neurônios por camada, também encontrarmos um comportamento oposto ao esperado. Esperava-se que, para um desvio padrão menor, tivéssemos um erro menor; porém, pelo apresentado, quanto maior o desvio padrão, menor é a somatória dos erros quadráticos. Entende-se que por uma rede neural possuir uma característica mais uniforme, a informação que circula em seu interior possua maior qualidade. O número de inflexões possui um coeficiente negativo e, portanto, significando que, quanto maior o número de inflexões, menor será erro. Novamente um resultado que vai em desencontro com a expectativa com resultado obtido a partir da análise anterior no item 0 das *top 40* arquiteturas. O $R^2$ ajustado foi de 0,0267, significando que as características da rede neural explicam relativamente pouco o erro cometido na previsão do retorno do ativo. Entende-se que por se tratar de uma relação não linear, de fato o $R^2$ ajustado será baixo. É possível perceber que, de fato, há relações não lineares quando, por exemplo realizamos a regressão linear utilizando o quadrado da variável explicativa. Para demonstrar este conceito, iremos utilizar a variável número de neurônios (N_NEURONIOS) e elevaremos ela ao quadrado para que seja realizado a regressão linear apenas olhando esta variável. Na Tabela 12, é possível perceber que é de fato significativo as variáveis explicativas N_NEURONIOS e N_NEURONIOS_2 (que nada mais é que o número de neurônios ao quadrado) e, portanto, provavelmente a relação do número de neurônios com o erro cometido pela rede neural é não linear. Embora não seja apresentado aqui, podemos realizar a mesma inferência para as outras características, o que seria uma das explicações para o $R^2$ ajustado ser baixo. Logicamente, existem outras variáveis que influenciam no erro de forma até mais relevante, porém estas não participarão da análise da RLM.

Tabela 12 - Regressão linear do erro pelas características

| Variável explicativa | Coeficiente | Desvio Padrão | t | p-valor>|t| | Intervalo com 95% de confiança |
|---|---|---|---|---|---|
| N_NEURONIOS | -0,0007 | 0,0003 | -2,17 | 0,0300 | [-0,0012958, -0,0000651] |
| N_NEURONIOS_2 | 0,0000 | 0,0000 | 4,12 | 0,0000 | [0,0000201, 0,0000567] |
| _cons | 0,2348 | 0,0026 | 91,15 | 0,0000 | [0,2297912, 0,2398922] |

Percebemos aqui uma forte influência do número de camadas, do número de neurônios e do número de inflexões. E diferente do que se imaginava o número de inflexões apresenta melhores informações do que o desvio padrão do número de neurônios entre as camadas. Outra importante informação é o erro cometido pelas redes neurais, que está na ordem de 0,2%, ou seja, dado que o módulo do retorno é da ordem de 2%, o erro é significativamente pequeno.

# 6 Conclusão

Este trabalho procurou explicar como as diferentes características de arquitetura de redes neurais afetam seu desempenho em termos de assertividade. Como resultado, desenvolvemos um estudo que utiliza o método exaustivo conhecido também como *força bruta,* para criar uma população de arquiteturas de redes neurais com seus diferentes desempenhos. Desta população extraímos as melhores redes neurais em termos de assertividade e apresentamos, de forma estatística, quais as características que as diferenciaram. Por se tratar de um problema não linear, como corroborado nos resultados das regressões lineares, torna-se difícil estabelecer uma correlação entre uma característica de arquitetura de rede e sua assertividade. É interessante observar que os coeficientes apresentados pela regressão tiveram o sinal de acordo com o esperado para boa parte das características. Em alguns casos, como foi visto para o número de neurônios, quanto maior o número maior o erro. Porém vale ressaltar que, conforme foi apresentado, não se trata de uma relação linear. Um resultado interessante que se mostrou bastante consistente foi possível verificar a existência de um intervalo de número de neurônios em que as redes neurais possuem desempenho assertivo pior. Este intervalo pode variar de acordo com diversas características da arquitetura de uma rede neural e até a forma como o método exaustivo é parametrizado. Percebemos que para uma determinada população de redes neurais que busca a resolução de um determinado problema, podemos ter características que influenciam diretamente no desempenho. Apresentamos neste trabalho também uma característica não presente em outros estudos que é o número de inflexões. Como mostrado no item 5.1, é possível perceber que se destacam negativamente arquiteturas que possuem maior número de inflexões, e para alguns casos, quando não há inflexões, o desempenho aparentemente foi melhor, fazendo com que aquela arquitetura estivesse entre as *top 40*. Consegue-se a partir das observações dos erros absolutos, perceber que as melhores arquiteturas possuem assertividades muito diferentes entre si. Portanto, concluímos que de fato, a arquitetura influencia no desempenho de uma rede neural, mostrando-se um fator de grande impacto na sua assertividade. O estudo teve como limitação o poder de processamento do equipamento utilizado, o que restringiu a população de redes neurais geradas. Além disso, o problema de previsão de retorno de ações possui limitações devido às muitas variáveis que o influenciam, sendo a correlação destas variáveis não linear entre si. A forma como o a metodologia é aplicada pode indicar um caminho importante para o desenvolvimento de modelos de otimização de treinamento, tais como o proposto por Falhman (1989). Outra aplicação seria a otimização de algoritmos tais como o de seleção genética, pois trataria a geração de uma população de forma orientada e não totalmente aleatória. Ainda é possível expandir o espaço amostral usado para estudo aumentando a assertividade do trabalho, utilizando-se de um poder de processamento maior e uma série de dados mais abrangente este problema poderia ser cercado com mais assertividade. Além disso considerar outros fatores que influência na arquitetura tais como funções sinápticas e técnica de treinamento aplicada. A forma como as diferentes características são analisadas pode ser insuficiente dado que a relação entre as características das redes neurais e sua assertividade para previsão podem não ser linear. Dessa forma, outros métodos podem se mostrar mais promissores para provar estas correlações.

## 7    Referências Bibliográficas